\documentclass[aps,preprintnumbers,onecolumn,amsmath,amssymb,floatfix,pra]{revtex4} 
\pdfoutput=1
\usepackage[utf8]{inputenc}
\usepackage{amsmath}
\usepackage{amsfonts}
\usepackage{braket}
\usepackage{tensor}
\usepackage{bbold}
\usepackage[colorlinks=true,linkcolor=blue, citecolor=blue, bookmarks]{hyperref}
\def\be{\begin{equation}}
\def\ee{\end{equation}}
\def\bea{\begin{eqnarray}}
\def\eea{\end{eqnarray}}

\begin{document}

\title{Recursive formulas for the overlaps between Bethe states and product states in XXZ Heisenberg chains}

\author{Lorenzo Piroli$^{1,2}$ and Pasquale Calabrese$^1$ }

\affiliation{
$^1$ Dipartimento di Fisica dell'Universit\`a di Pisa and INFN, 56127 Pisa, Italy;\\
$^2$ Scuola Normale Superiore, Piazza dei Cavalieri 7, 56126 Pisa, Italy. 
}

\begin{abstract}

We consider the problem of computing the overlaps between the Bethe states of the XXZ spin-1/2 chain and generic states. 
We derive recursive formulas for the overlaps between some simple product states and off-shell Bethe states within 
the framework of the Algebraic Bethe Ansatz. 
These recursive formulas can be used to prove in a simple and straightforward way the recently-obtained results for the overlaps 
of the Bethe states with the N\'eel state, the dimer state, and the \textit{q}-deformed dimer state.
However, these recursive formulas are derived for a broader class of states
and represent a concrete starting point for the computation of rather general overlaps.
Our approach can be easily extended to other one-dimensional Bethe Ansatz integrable models.

\end{abstract}

\maketitle

\section{Introduction}\label{intro}
The last decade has witnessed an increasing theoretical and experimental effort in the study of the out-of-equilibrium dynamics 
of isolated many-body quantum systems. 
In particular, the  time evolution following a sudden change in one of the parameters of the Hamiltonian (quantum quench) has received 
a lot of attention \cite{silva}. 
A number of theoretical and experimental investigations have unambiguously shown that for large times after a quantum quench
and in the thermodynamic limit, the expectation values of the local observables relax to stationary values, 
although the dynamics governing the evolution is unitary. 
For a generic system it has been argued and shown in a series of numerical and experimental works that these 
stationary values are described by a Gibbs distribution in which the (effective) temperature is fixed 
by the energy of the initial state \cite{nonint,rdo-08,tvar,tetal-11}.
Oppositely, integrable systems  keep memory of many details of the initial state also for infinite time \cite{kww-06,gg,rs-12,ce-13}, 
as a consequence of the constrained dynamics with an infinite number of (local) conservation laws. 
It has been conjectured that, for an integrable system, these stationary values can be calculated  using a 
generalised Gibbs ensemble (GGE), a statistical ensemble determined by all local conserved charges \cite{gg}
(the importance and the role of locality of the integrals of motion have been highlighted 
mainly in Refs. \cite{calabrese_essler_fagotti_II,fe-13}).
A small perturbation close to  integrability leads to interesting pre-thermalisation effects which are captured by a 
(perturbed) GGE \cite{ekmn-14,mmgs-13}.

Many analytical works have focused on providing exact results to test the GGE predictions in specific many-body integrable models. 
The predictions based on GGE resisted all tests in many models both having a free-particle representation
\cite{fe-13,cazalilla,cdeo-08,bs-08, cef,calabrese_essler_fagotti_II,eef-12,de-12,ck-14,mc-12,csc13,sc-14,bkc-14} and being genuinely 
interacting \cite{ksc-13,kcc14, de_nardis_wouters,ck-12,mckc-14}, until very recently when it has been found that for the XXZ spin chain, 
after a quench from the (symmetrized) N\'eel and dimer states, the obtained stationary 
values  \cite{pozsgay_werner_kormos, wouters_brockmann,noGGE4}
disagree with the predictions of the GGE built with all the known local charges \cite{fe-13b,fcce-13} 
(see also Refs. \cite{f-14,noGGEp,noGGE3}).
It is worth stressing that, besides representing a test for the validity of the GGE, exact results for the time evolution of local observables 
are also extremely useful to gain insight into the relaxation dynamics. 

Le us briefly recall what are the needed building blocks to study the quench dynamics in a generic situation. 
The first problem one faces is to write the initial state $|\Phi_0\rangle$ in terms of the eigenstates of the Hamiltonian $H$ governing 
the time evolution. 
Let us for the moment generically denote the {\it normalised} eigenstates as $|n\rangle$, in such way that the initial state can be written as
\be
|\Phi_0\rangle=\sum_n a_n |n\rangle,
\ee
where $a_n$ are the {\it overlaps} $a_n\equiv \langle n|\Phi_0\rangle$ between the initial state and the 
eigenstates. Consequently, the time evolved state is 
\be
|\Phi(t)\rangle=\sum_n a_n e^{-i E_n t}|n\rangle,
\ee
where $E_n$ is the energy of the state $|n\rangle$.
This provides the time dependent expectation values of an arbitrary observable $O$, 
in terms of the form factors $\langle n|O|m\rangle$, as
\be
\langle \Phi(t)|O |\Phi(t)\rangle= \sum_{mn} a_n a^*_m e^{-i(E_n- E_m) t} \langle m| O |n\rangle\,.
\label{sumff}
\ee
Summing up, in order to  characterise the quench dynamics, the needed starting elements are 
(i) a complete characterisation of all eigenstates $|n\rangle$ of a Hamiltonian and their energies; 
(ii) the norms of the eigenstates and the form factors of relevant operators in this basis;
(iii) the overlaps between the initial states and the eigenstates. 

For integrable models, the Bethe Ansatz is a very efficient tool to obtain most of these ingredients. 
Indeed, it provides a full set of eigenstates with their energies \cite{gaudin}.
The norms and the form factors of the most relevant local operators are the main objectives 
of the Algebraic Bethe Ansatz and quantum inverse scattering program \cite{korepin_book}.
What is not (yet) known in general is how to obtain the overlaps between Bethe states and 
generic initial states.
Up to now, very few exact results exist for these overlaps in integrable 
models \cite{faribault_calabrese_caux, gritsev, pozsgay_l_echo, pozsgay, brockmann_I, brockmann_II, brockmann_III,cd-14}. 
Clearly, finding compact and tractable expressions for the overlaps between Bethe states and the initial state in a 
generic quantum quench would allow exact calculations for a variety of potentially interesting situations and experiments. 

However, the ingredients listed above are only the starting point for the description of the quench dynamics,
because the sum (\ref{sumff}) is still to be performed and this is a very difficult step. 
Indeed, the same problem is also present in the calculation of the equilibrium correlation functions for which 
the knowledge of the form factors \cite{kks-97,kmt}  allows analytical calculations only in a few instances/limits 
(see e.g. Refs. \cite{kmt2, kk, gohmann, pozsgay_correlations} for the XXZ chain but this list is far from being exhaustive). 
Accurate determinations of the equilibrium correlation functions can be obtained by summing the form factor expansion numerically for 
finite systems, as e.g. done for the XXZ chain \cite{cm-05} and for the Lieb-Liniger model \cite{cc-06g}.
For what concerns the non-equilibrium quench dynamics, the summation problem is still present even after knowing the overlaps.
Fortunately, a recently-proposed method (termed either representative state approach or quench action) gives an exact analytical 
description of the post-quench steady state in the thermodynamic limit \cite{ce-13}.  
The essential building blocks of this method are once again the overlaps.
Thus the main obstacle to tackle quite generally the quench dynamics is to find compact and manageable expressions for 
the overlaps which could be subsequently used both for numerical and analytical calculations 
(for the sake of completeness, we must mention that there are also some approaches for studying quench dynamics in 
integrable models partially by-passing the calculations of the overlaps such as imaginary time formalism \cite{cc-06,fm-10,gc-11,sfm-12,c-14}, 
Yudson representation \cite{a-12,wa-14}, and others \cite{cro,jap,m-13,bse-14}).

In this work we derive recursive formulas for the overlaps between Bethe states in the XXZ spin-1/2 chain and a class of product states. 
The structure of the paper is as follows. In Section \ref{aba} we review the XXZ model and the Algebraic Bethe Ansatz 
tools that will be used in the rest of this work. 
In Section \ref{recursive_formulas} we introduce a certain class of product states and we derive recursive formulas for the overlaps 
between these states and Bethe states. 
The class we consider includes the N\'eel state, the dimer state, and the $q$-deformed dimer state, for which overlaps formulas 
were recently derived by B. Pozsgay using Boundary Bethe Ansatz techniques \cite{pozsgay}.
In Section \ref{determinant_formulas} we show that our recursive formulas can be used to prove in a simple way the overlaps 
of Ref. \cite{pozsgay}. 
Finally, conclusions are presented in Section \ref{conclusions}.

\section{The XXZ spin chain and the Algebraic Bethe Ansatz}\label{aba}

We consider  the XXZ spin-1/2 chain with Hamiltonian
\begin{equation}
H^{XXZ}_{N}=
\sum_{j=1}^{N}\left[\sigma^{x}_{j}\sigma^{x}_{j+1}+\sigma^{y}_{j}\sigma^{y}_{j+1}+\Delta\left(\sigma^{z}_{j}\sigma^{z}_{j+1}-1\right)\right]\, ,
\label{eq:1}
\end{equation}
where $\sigma^{\alpha}_{j}$ are the Pauli matrices at the site $j$ and we impose
periodic boundary conditions $\sigma^{\alpha}_{N+1}=\sigma^{\alpha}_{1}$. 
The Hamiltonian \eqref{eq:1} is defined in the Hilbert space $\mathcal{H}_{N\ldots 1}=h_N\otimes\ldots\otimes h_1$, 
where $h_i\simeq \mathbb{C}_2$ is associated with site $i$ of the chain.

The Hamiltonian \eqref{eq:1} commutes with the $z$-component of the total spin so that the Hilbert space 
$\mathcal{H}_{N\ldots 1}$ can be decomposed into sectors with a well-defined number $P$ of flipped spins 
with respect to the reference state with all spins up
\begin{equation}
\ket{0}_{N\ldots 1}=\ket{\uparrow}_N\otimes\ldots \otimes\ket{\uparrow}_1 .
\label{eq:8}
\end{equation}
The wave-function in the spin basis and in the sector with $P$ spins down is given by the ansatz \cite{gaudin}
\be
\Psi(\lambda_1,\dots \lambda_P| s_1,\dots s_P)= \sum_{Q\in \sigma_P} \prod_{j=1}^P F(\lambda_{Q_j},s_j)
\prod_{k<j}\frac{\sinh(\lambda_{Q_j}-\lambda_{Q_k}-\eta)}{\sinh(\lambda_{Q_j}-\lambda_{Q_k})}\,,
\label{psi}
\ee
where $s_j$ denotes the position of the down spins (we assumed, without loss of generality, $s_j<s_k$ for $j<k$), 
and we introduced $\eta={\rm arccosh}\, \Delta$ and the function 
\be
F(\lambda,s) = \sinh(\eta)\sinh^{s-1}(\lambda+\eta/2)\sinh^{2N-s}(\lambda-\eta/2).
\ee
The sum runs over the permutations $Q$ of $P$ elements.
The wave function (\ref{psi}) is an eigenstate of the Hamiltonian (\ref{eq:1}) with periodic boundary conditions  
if the {\it rapidities} $\lambda_j$ satisfy the Bethe equations \cite{gaudin}
\begin{equation}
\left(\frac{\sinh(\lambda_j-\eta/2)}{\sinh(\lambda_j+\eta/2)}\right)^{N}=\prod_{\substack{l=1\\l\neq j}}^{P}\frac{\sinh(\lambda_j-\lambda_l-\eta)}{\sinh(\lambda_j-\lambda_l+\eta)}\ ,
\label{eq:9}
\end{equation}
and the corresponding energies are
\begin{equation}
E\left(\{\lambda_j\}\right)=\sum_{j=1}^{P}\frac{4\sinh^{2}(\eta)}{\cosh(2\lambda_{j})-\cosh(\eta)}\ .
\label{eq:11}
\end{equation}
In the thermodynamic limit $N,P\to \infty$ with $P/N$ constant, the 
properties of the XXZ spin-chain can be obtained by means of the Thermodynamic Bethe Ansatz \cite{tba}.
In particular it turns out that the model is gapped for $\Delta>1$ and gapless in the opposite regime. 

The wave function (\ref{psi}) gives already an explicit form for the overlaps with product states, 
as a sum over the permutations of the $P$ elements.
However, as stressed already elsewhere \cite{pozsgay}, since the number of permutations grows like $P!$, these expressions 
for the overlaps are not useful for any practical numerical or analytic evaluation.

A powerful alternative for the study of the XXZ chain (and in general of Bethe Anstaz solvable models) is the 
Algebraic Bethe Ansatz that we are are going to briefly review now,  
remanding for a more detailed treatment to the standard textbook on the subject \cite{korepin_book}.  
The central object of the Algebraic Bethe Ansatz is the $R$-matrix which, for the XXZ model, is 
(here and below all the non written elements of matrices are equal to zero)
\begin{equation}
R(\lambda,\mu)=\begin{pmatrix}
  f(\mu,\lambda) &  &  & \\
   & g(\mu,\lambda) &1  &  \\
   & 1 &  g(\mu,\lambda)& \\
   &  & & f(\mu,\lambda)
 \end{pmatrix} \ ,
 \label{eq:2}
\end{equation}
where
\begin{equation}
f(\mu,\lambda)=\frac{\sinh(\lambda-\mu+\eta)}{\sinh(\lambda-\mu)}\ , \qquad g(\mu,\lambda)= \frac{\sinh(\eta)}{\sinh(\lambda-\mu)}\  .
\label{eq:3}
\end{equation} 
An auxiliary space $h_0\simeq \mathbb{C}_2$ is introduced together with the $L$-operator acting on 
the four-dimensional space $h_0\otimes h_n$
\begin{equation}
L_{0,n}(\lambda)=\begin{pmatrix}
  \sinh(\lambda+\frac{\eta}{2}) &  &  & \\
   & \sinh(\lambda-\frac{\eta}{2}) &\sinh(\eta)  &  \\
   & \sinh(\eta) &  \sinh(\lambda-\frac{\eta}{2})& \\
   &  & & \sinh(\lambda+\frac{\eta}{2})
 \end{pmatrix}\ .
\label{eq:4}
\end{equation}
A second fundamental object of the Algebraic Bethe Ansatz, the monodromy matrix, is then defined as the product of $L$-matrices along the chain
\begin{equation}
T_{N \ldots 1}(\lambda)=L_N(\lambda)\ldots L_1(\lambda)=\begin{pmatrix}
  A_{N \ldots 1}(\lambda) & B_{N \ldots 1}(\lambda) \\
  C_{N \ldots 1}(\lambda) & D_{N \ldots 1}(\lambda)\\
 \end{pmatrix} \ ,
\label{eq:5}
\end{equation}
where we have dropped the index corresponding to the auxiliary space. The entries of the monodromy matrix are operators acting on $\mathcal{H}_{N\ldots 1}$. The $R$-matrix \eqref{eq:2} and the monodromy matrix \eqref{eq:5} fulfil  the following fundamental relation
\begin{equation}
R(\lambda,\mu)\Big(T_{N \ldots 1}(\lambda)\otimes T_{N \ldots 1}(\mu)\Big)=\Big(T_{N \ldots 1}(\mu)\otimes T_{N \ldots 1}(\lambda)\Big)R(\lambda,\mu)\ .
\label{eq:6}
\end{equation}
Writing down explicitly equation \eqref{eq:6} results in a number of quadratic relations for the entries of the monodromy matrix (see \cite{korepin_book} for the details).

The reference state (\ref{eq:8}) is an eigenstate of the diagonal entries of the monodromy matrix, indeed it holds
\begin{align}
&A_{N \ldots 1}(\lambda)\ket{0}_{N\ldots 1}=a_{N \ldots 1}(\lambda)\ket{0}_{N \ldots 1}=\sinh^{N}(\lambda+\frac{\eta}{2})\ket{0}_{N \ldots 1}\ ,\label{eq:8.1}\\
&D_{N \ldots 1}(\lambda)\ket{0}_{N\ldots 1}=d_{N \ldots 1}(\lambda)\ket{0}_{N \ldots 1}=\sinh^{N}(\lambda-\frac{\eta}{2})\ket{0}_{N \ldots 1}\ .\label{eq:8.2}
\end{align}
In this algebraic formalism, the eigenstates of Hamiltonian \eqref{eq:1} (lying in the $P$-sector)
are then written in terms of the entries of the monodromy matrix as
\begin{equation}
\ket{\{\lambda_j\}_{j=1}^{P}}=B(\lambda_P)\ldots B(\lambda_1)\ket{0}_{N\ldots 1}\ ,
\label{eq:10}
\end{equation}
where, once again, the rapidities $\{\lambda_j\}_{j=1}^P$ fulfil the Bethe equations (\ref{eq:9}) showing that Coordinate and 
Algebraic Bethe Ansatz are indeed equivalent. 

A crucial property of this algebraic construction is that the state \eqref{eq:10} is well defined even if the 
parameters $\{\lambda_j\}_{j=1}^{P}$ do not fulfil the Bethe equations \eqref{eq:9} (but obviously is not an eigenstate 
of the Hamiltonian (\ref{eq:1})).
Among the most remarkable results obtained by means of the Algebraic Bethe Ansatz, we must mention 
the proof of the Gaudin formula for the norm of a Bethe state \cite{korepin_norm}, the Slavnov formula \cite{slavnov} which is a a
general expression for the scalar product between two Bethe states of the form (\ref{eq:10}) in which 
only one of the two sets of rapidities satisfies the Bethe equations, and the determination of the form factors 
of the most relevant operators \cite{kmt,kk,kks-97}.

\section{Derivation of recursive formulas}\label{recursive_formulas}
In this section we derive recursive formulas for the overlaps between Bethe states and a class of product states. 
Our approach makes use of a two-sites generalised model \cite{korepin_book, mossel_caux} and is based on a particular representation 
of the Bethe states in this model. 
In subsection \ref{derivation} we introduce the class of product states considered in this work and the two-sites generalised model 
representation for Bethe states. 
General recursive formulas are then derived. In subsection \ref{examples} we give explicit examples for a number 
of physically relevant states.
 
\subsection{Two-sites generalized model}\label{derivation}
Suppose the number of sites of the chain $N$ is divisible by the integer $G$. In this work we consider product states of the following form
\begin{equation}
\ket{\psi}_{N\ldots 1}=\ket{\varphi}_{N,N-1,\ldots, N-G+1}\otimes\ket{\varphi}_{N-G,N-G-1,\ldots, N-2G+1}\otimes\ldots\otimes\ket{\varphi}_{G,G-1,\ldots,1}\ ,
\label{eq:12}
\end{equation}
where the state $\ket{\varphi}_{r, r-1, \ldots, r-G+1}$ belongs to the Hilbert space $h_{r}\otimes h_{r-1}\otimes\ldots \otimes h_{r-G+1}$.
These states include and generalise those considered in Ref. \cite{pozsgay} for $G=2$ and those in Ref. \cite{fcce-13} 
for $G=1,2$.

We are interested in deriving recursive formulas for the overlaps between states of the form \eqref{eq:12} and Bethe states
(\ref{eq:10}). We start considering a chain with $N+G=G(M+1)$ sites. The scalar product we are interested in is
\begin{equation}
S_{M+1}[\psi](\lambda_1, \ldots ,\lambda_P)= \tensor[_{G(M+1) \ldots 1}]{\bra{\psi}}{}B(\lambda_P)\ldots B(\lambda_1)\ket{0}_{G(M+1) \ldots 1}\ .
\label{eq:13}
\end{equation}
The two-site generalised  model of Ref. \cite{korepin_book} allows us to write the monodromy matrix as follows
\begin{equation}
\begin{split}
T_{N+G \ldots 1}(\lambda)&=L_{N+G}(\lambda)\ldots L_{N+1}(\lambda)T_{N \ldots 1}(\lambda)=T_{N+G\ldots N+1}(\lambda)T_{N \ldots 1}(\lambda)=\\
&=\begin{pmatrix}
  A_{N+G\ldots N+1}(\lambda) & B_{N+G\ldots N+1}(\lambda)  \\
  C_{N+G\ldots N+1}(\lambda) & D_{N+G\ldots N+1}(\lambda)
 \end{pmatrix}\begin{pmatrix}
  A_{N \ldots 1}(\lambda) & B_{N \ldots 1}(\lambda)  \\
  C_{N \ldots 1}(\lambda) & D_{N \ldots 1}(\lambda)
 \end{pmatrix}\ ,
\end{split}
\label{eq:14}
\end{equation}
where the operators $ A_{N+G\ldots N+1}(\lambda)$, $B_{N+G\ldots N+1}(\lambda)$, $C_{N+G\ldots N+1}(\lambda)$ and $D_{N+G\ldots N+1}(\lambda)$ act on the Hilbert space $h_{N+G}\otimes h_{N+G-1}\ldots \otimes h_{N+1}$, while $A_{N \cdots 1}(\lambda)$,$B_{N \cdots 1}(\lambda)$, $C_{N \ldots 1}(\lambda)$ and $D_{N \ldots 1}(\lambda)$ act on $\mathcal{H}_{N \ldots 1}$. 
For convenience, we introduce the following notations for the operators and the reference states
\begin{align}
&\widetilde{X}(\lambda)\equiv X_{N+G\ldots N+1}(\lambda)\ ,
\label{eq:15}\\
&X(\lambda)\equiv X_{N \ldots 1}(\lambda)\ ,
\label{eq:16}\\
&\ket{\widetilde{0}}\equiv \ket{\uparrow}_{N+G}\otimes \ket{\uparrow}_{N+G-1}\otimes\ldots\otimes\ket{\uparrow}_{N+1}, \label{eq:17}\\
&\ket{0}\equiv \ket{0}_{N \ldots 1}=\ket{\uparrow}_N \otimes \ldots \otimes \ket{\uparrow}_1\ .
\label{eq:18}
\end{align}
Note that the reference state of the whole chain is the product of two reference states, $\ket{0}_{N+G \ldots 1}=\ket{\widetilde{0}} \otimes \ket{0}$. From Eq. \eqref{eq:14} we have
\begin{equation}
B_{N+G \ldots 1}(\lambda)= \widetilde{A}(\lambda)B(\lambda)+\widetilde{B}(\lambda)D(\lambda)\ ,
\label{eq:19}
\end{equation}
and, using Eqs.  \eqref{eq:8.1} and \eqref{eq:8.2}, the following relations hold 
\bea
\widetilde{A}(\lambda)\ket{\widetilde{0}}&=&
\widetilde{a}(\lambda)\ket{\widetilde{0}}=\sinh^G\left(\lambda+\frac{\eta}{2}\right)\ket{\widetilde{0}} \ ,
\label{eq:20}\\
D(\lambda)\ket{0}&=&
d(\lambda)\ket{0}=\sinh^N\left(\lambda -\frac{\eta}{2}\right)\ket{0}=\sinh^{GM}\left(\lambda -\frac{\eta}{2}\right)\ket{0}\ .
\label{eq:21}
\eea
We can now write the Bethe state in Eq. \eqref{eq:13} as
\be
B(\lambda_{P}) \ldots B(\lambda_1) \ket{0}_{N+G \ldots 1} = \Big[\left(\widetilde{A}(\lambda_{P})B(\lambda_{P})+\widetilde{B}(\lambda_{P})D(\lambda_{P})\right) \cdots
\left(\widetilde{A}(\lambda_1)B(\lambda_1)+\widetilde{B}(\lambda_1)D(\lambda_1)\right)\Big] \ket{\widetilde{0}}\otimes \ket{0}\ .
\label{eq:22}
\ee
The key representation for Bethe states is obtained expanding the product in \eqref{eq:22} and collecting together all terms 
with the same number of $\widetilde{B}$ operators, arriving finally to
\begin{equation}
B(\lambda_{P}) \ldots B(\lambda_1) \ket{0}_{N+G \ldots 1} =\sum_{J=0}^{P}\ket{\Gamma_J} \ ,
\label{eq:23}
\end{equation}
where $\ket{\Gamma_J}$ contains all the terms with a number $J$ of $\widetilde{B}$ operators that result from expanding the 
product in Eq. \eqref{eq:22}. 
The explicit expression for $\ket{\Gamma_J}$ can be found in \cite{korepin_book}:
\be
\ket{\Gamma_J}=\sum_{\substack{\{\lambda\}=\{\lambda^{I}\}\cup \{\lambda^{II}\}\\|\{\lambda^I\}|=J}}
\left(\prod_{l=1}^{J}\prod_{m=1}^{P-J}\widetilde{a}(\lambda_{m}^{II})d(\lambda_l^{I}) f(\lambda^{II}_{m}, \lambda^{I}_l)\right)
\prod_{r=1}^{J}\widetilde{B}(\lambda^{I}_{r})\ket{\widetilde{0}}\otimes \prod_{s=1}^{P-J} B(\lambda^{II}_{s})\ket{0} \ ,
\label{eq:24}
\ee
where the summation is over all decomposition of $\{\lambda_{j}\}_{j=1}^{P}$ into two disjoint subsets $\{\lambda^{I}_{j}\}$ 
and $\{\lambda^{II}_{j}\}$, such that the cardinality of the set $\{\lambda^{I}_{j}\}$ is equal to $J$. As an example, we give the 
explicit expression of $\ket{\Gamma_J}$ for $J=0,1,2$
\begin{align}
&\ket{\Gamma_0}=\prod_{j=1}^{P}\widetilde{a}(\lambda_j)\ket{\widetilde{0}}\otimes\prod_{l=1}^{P}B(\lambda_l)\ket{0}\ ,\label{eq:25}\\
&\ket{\Gamma_1}=\sum_{j=1}^{P}d(\lambda_j)\prod_{\substack{k=1 \\ k\neq j}}^{P} \widetilde{a}(\lambda_k) f(\lambda_k,\lambda_j) \widetilde{B}(\lambda_j)\ket{\widetilde{0}}\otimes  \prod_{\substack{l=1\\ l\neq j}}^{P} B(\lambda_l) \ket{0} \ ,\label{eq:26} \\
&\ket{\Gamma_2}=\sum_{\substack{l,j\\l<j}}^{P}d(\lambda_j)d(\lambda_l)
\Big(\prod_{\substack{k=1 \\ k\neq l, j}}^{P} \widetilde{a}(\lambda_k) f(\lambda_k,\lambda_l)f(\lambda_k,\lambda_j) \Big)
 \widetilde{B}(\lambda_j)\widetilde{B}(\lambda_l)\ket{\widetilde{0}}\otimes  \prod_{\substack{r=1\\ r\neq l, j}}^{P} B(\lambda_r) \ket{0} \ .
\label{eq:27}
\end{align}

We now take the scalar product between the Bethe state \eqref{eq:23} with the product state
\begin{equation}
\ket{\psi}_{N+G\ldots 1}=\ket{\varphi}_{N+G\ldots N+1}\otimes \ket{\psi}_{N\ldots 1}\ .
\label{eq:27.1}
\end{equation}
Using the same notation as in Eq. \eqref{eq:13} for the scalar product, 
from Eqs. \eqref{eq:23}, \eqref{eq:24}, and \eqref{eq:27.1}, we finally arrive at the recursive formula
\begin{equation}
\begin{split}
&S_{M+1}[\psi](\lambda_1, \ldots ,\lambda_P)=\\
&= \sum_{J=0}^{P}\ \sum_{\substack{\{\lambda\}=\{\lambda^{I}\}\cup \{\lambda^{II}\}\\|\{\lambda^I\}|=J}}\left(\prod_{l=1}^{J}\prod_{m=1}^{P-J}\widetilde{a}(\lambda_{m}^{II})d(\lambda_l^{I}) f(\lambda^{II}_{m}, \lambda^{I}_l)\right) \braket{\varphi|\prod_{r=1}^{J}\widetilde{B}(\lambda^{I}_{r})|\widetilde{0}} 
S_{M}[\psi](\lambda^{II}_1, \ldots ,\lambda^{II}_{P-J})\ .
\end{split}
\label{eq:28}
\end{equation}
The  quantity $\braket{\varphi|\prod_{l=1}^{J}\widetilde{B}(\lambda^{I}_{l})|\widetilde{0}}$ can be computed using Eq. \eqref{eq:17} 
and the definition of the $\widetilde{B}$ operator given in Eq. \eqref{eq:14}.
In order to derive the general expression \eqref{eq:28}, we have considered all the terms $\ket{\Gamma_J}$ in the sum in Eq. \eqref{eq:23}. 
On the other hand, for specific examples of states of the form \eqref{eq:12}, only few terms in the sum in Eq. \eqref{eq:23} 
have to be considered, because most of the vectors $\ket{\Gamma_J}$ have zero overlap with them.

Specifically, the number of vectors $\ket{\Gamma_J}$ in Eq. \eqref{eq:23} having nonzero overlap with the product 
state \eqref{eq:12} is related to the number of down spins of $\ket{\varphi}$ in Eq. \eqref{eq:12}. 
This is because $\ket{\Gamma_J}$ is the sum of terms containing a number $J$ of $\widetilde{B}$ operators and thus lies in the 
$J$-sector of the Hilbert space $h_{N+G}\otimes \ldots \otimes h_{N+1}$. 
For example, if $\ket{\varphi}$ lies in the one-sector of the corresponding Hilbert space, the only term in the sum \eqref{eq:23} 
with nonzero overlap is $\ket{\Gamma_1}$. 
Accordingly, when specific states are considered, Eq. \eqref{eq:28} can be greatly simplified to obtain tractable expressions.

Note that given the integer $G$ in Eq. \eqref{eq:12}, the $\widetilde{B}$ operator is obtained as the entry of the product of 
$G$ consecutive $L$-matrices, according to Eq. \eqref{eq:14}. 
For increasing values of $G$, more lengthy calculations are thus needed for the computation of 
$\braket{\varphi|\prod_{r=1}^{J}\widetilde{B}(\lambda^{I}_{r})|\widetilde{0}}$ which is necessary for deriving explicit expressions for 
the recursive formulas of specific states.

In the next subsection we consider a number of physically relevant states and present the corresponding recursive formulas for the overlaps with Bethe states.

\subsection{Recursive formulas for specific states}\label{examples}
We now write explicitly the general recursive formula \eqref{eq:28} for the following specific physically relevant states of the form \eqref{eq:12}.  
\begin{enumerate}
\item The ferromagnet along the $x$-direction, $\ket{xF}=\ket{\rightarrow\ldots\rightarrow}$. The state $\ket{xF}$ is of the form \eqref{eq:12} with $G=1$ and $\ket{\varphi}=\ket{\rightarrow}=(\ket{\uparrow}+\ket{\downarrow})/\sqrt{2}$. In this case, we only have to consider $\ket{\Gamma_0}$ and $\ket{\Gamma_1}$ in the expansion \eqref{eq:23}, the other terms having zero overlap as discussed in the previous subsection. 
Since $G=1$, the definition of $\widetilde{B}$ involves only one $L$-matrix. From Eq. (\ref{eq:4}) we have directly
\begin{equation}
\widetilde{B}(\lambda_j)=\begin{pmatrix}
0&0\\
\sinh(\eta)&0
\end{pmatrix} \ ,
\label{a}
\end{equation}
so that
\begin{equation}
\braket{\rightarrow|\widetilde{B}(\lambda_j)|\uparrow}=\begin{pmatrix}\frac{1}{\sqrt{2}}&\frac{1}{\sqrt{2}}\end{pmatrix}
\begin{pmatrix}
0&0 \\
\sinh(\eta)&0
\end{pmatrix} \begin{pmatrix}1\\0 \end{pmatrix}=\frac{\sinh(\eta)}{\sqrt{2}} \ .
\label{bb}
\end{equation}

The recursive formula \eqref{eq:28} can thus be simplified as follows
\begin{equation}
\begin{split}
& S_{N+1}[xF](\lambda_1 , \ldots , \lambda_P)= \frac{1}{\sqrt{2}}\left[ \prod_{k=1}^{P}\sinh\left( \lambda_k+\frac{\eta}{2}\right)\right] S_N[xF](\lambda_1 , \ldots , \lambda_P) +\\ 
& \hspace{2cm}+\frac{\sinh(\eta)}{\sqrt{2}} \sum_{j=1}^{P}\sinh^{N}\left(\lambda_j-\frac{\eta}{2}\right) \left[ \prod_{\substack{k=1\\ k\neq j}}^{P}\sinh\left( \lambda_k+\frac{\eta}{2}\right) f(\lambda_k,\lambda_j) \right]
S_N[xF](\lambda_1 , \ldots , \widehat{\lambda}_j , \ldots , \lambda_P)\ ,
\end{split}
\label{eq:30}
\end{equation}
where the notation $\hat{\lambda}_j$ means, as usual,  that the rapidity $\lambda_j$ is removed from the set 
$\{\lambda_1,\ldots, \lambda_P\}$ and $f(\mu,\nu)$ is given in Eq. (\ref{eq:3}).
\item The tilted ferromagnet $\ket{\theta F}= \ket{\theta;\nearrow,\ldots,\nearrow}=e^{i\theta/2\sum_j \sigma_j^y}\ket{\uparrow\ldots\uparrow}$. As before, we have $G=1$, but in this case $\ket{\varphi}=\cos(\theta/2)\ket{\uparrow}-\sin(\theta/2)\ket{\downarrow}$. 
Thus, analogously to Eq. (\ref{bb}), we simply have
\begin{equation}
\braket{\theta;\nearrow|\widetilde{B}(\lambda)|\uparrow}=-\sin(\theta/2)\sinh(\eta) \ ,
\label{eq:31}
\end{equation}
so that the corresponding recursive formula reads
\begin{equation}
\begin{split}
& S_{N+1}[\theta F](\lambda_1 , \ldots , \lambda_P)= \cos\left(\frac{\theta}{2}\right)\left[ \prod_{k=1}^{P}\sinh\left( \lambda_k+\frac{\eta}{2}\right)\right] S_N[\theta F](\lambda_1 , \ldots , \lambda_P) +\\
&\hspace{1cm} -\sin\left(\frac{\theta}{2}\right)\sinh(\eta) \sum_{j=1}^{P}\sinh^{N}\left(\lambda_j-\frac{\eta}{2}\right) \left[ \prod_{\substack{k=1\\ k\neq j}}^{P}\sinh\left( \lambda_k+\frac{\eta}{2}\right) f(\lambda_k,\lambda_j) \right]
S_N[\theta F](\lambda_1 , \ldots , \widehat{\lambda}_j , \ldots , \lambda_P)\ .
\end{split}
\label{eq:32}
\end{equation}
\item The N\'eel state $\ket{N}$,  the dimer state $\ket{D}$ and the $q$-deformed dimer state $\ket{qD}$ (considered in \cite{pozsgay})
\begin{align}
&\ket{N}=\ket{\uparrow\downarrow\uparrow\downarrow\ldots\uparrow\downarrow}=\otimes^{N/2}\ket{\uparrow\downarrow} \ ,\label{eq:33}\\
&\ket{D}=\otimes^{N/2}\frac{\ket{\uparrow\downarrow}-\ket{\downarrow\uparrow}}{\sqrt{2}}\ ,\label{eq:34}\\
&\ket{qD}=\otimes^{N/2}\frac{q^{1/2}\ket{\uparrow\downarrow}-q^{-1/2}\ket{\downarrow\uparrow}}{\sqrt{|q|+1/|q|}}, \qquad \Delta=(q+1/q)/2 \label{eq:35}\ .
\end{align}
The dimer state and the $q$-deformed dimer state are respectively the ground states of the Majumdar-Ghosh 
Hamiltonian \cite{majumdar} and the $q$-deformed Majumdar-Ghosh Hamiltonian \cite{batchelor}.

We now assume the number of sites $N$ to be even, $N=2M$. 
The states \eqref{eq:33}, \eqref{eq:34}, \eqref{eq:35} are of the form \eqref{eq:12} with $G=2$. 
The overlap between these states for $N=2M$ and a Bethe state is nonzero only if the number of rapidities corresponding to the 
Bethe state is $M$. 
Furthermore, in the computation of recursive formulas for these states, we only have to consider $\ket{\Gamma_1}$ in the 
expansion \eqref{eq:23}, the other terms having zero overlap. 
The recursive formulas for the three states \eqref{eq:33}, \eqref{eq:34}, \eqref{eq:35} thus read
\begin{equation}
\begin{split}
S_{M+1}&(\lambda_1, \ldots ,\lambda_{M+1})= \\
&= \sum_{j=1}^{M+1}\sinh^{2M}\left(\lambda_j-\frac{\eta}{2}\right)\braket{\varphi|\widetilde{B}(\lambda_j)|\uparrow\uparrow}
\left[ \prod_{\substack{k=1\\ k\neq j}}^{M+1}\sinh^2\left( \lambda_k+\frac{\eta}{2}\right) f(\lambda_k,\lambda_j) \right] S_{M}(\lambda_1 , \ldots , \hat{\lambda}_j , \ldots , \lambda_{M+1})\ .
\label{eq:36}
\end{split}
\end{equation}
The element $\braket{\varphi|\widetilde{B}(\lambda_j)|\uparrow\uparrow}$ can be simply obtained 
from the product of two $L$-matrices because in this case $G=2$. 
Using again Eq. (\ref{eq:4}) we have
\begin{equation}
\widetilde{B}(\lambda_j)= \begin{pmatrix}
\sinh(\lambda_j+\eta/2)&0 \\
0 &\sinh(\lambda_j-\eta/2)
\end{pmatrix}\otimes \begin{pmatrix}
0&0 \\
\sinh(\eta)&0
\end{pmatrix}+\begin{pmatrix}
0&0 \\
\sinh(\eta)&0
\end{pmatrix} \otimes \begin{pmatrix}
\sinh(\lambda_j-\eta/2)&0 \\
0&\sinh(\lambda_j+\eta/2)
\end{pmatrix}  ,
\label{c}
\end{equation}
so that for the N\'eel state
\begin{equation}
\begin{split}
&\braket{\uparrow\downarrow|\widetilde{B}(\lambda_j)|\uparrow\uparrow}=\\
&=\begin{pmatrix}1&0\end{pmatrix}\otimes\begin{pmatrix}0&1\end{pmatrix} \left[\begin{pmatrix}
\sinh(\lambda_j+\eta/2)&0 \\
0&\sinh(\lambda_j-\eta/2)
\end{pmatrix}\otimes \begin{pmatrix}
0&0 \\
\sinh(\eta)&0
\end{pmatrix}\right]\begin{pmatrix}1\\0 \end{pmatrix}\otimes \begin{pmatrix}1\\0 \end{pmatrix}=\sinh(\eta)\sinh(\lambda_j+\eta/2) .
 \label{eq:37}
\end{split}
\end{equation}
Analogously for the dimer state and the $q$-dimer state, we obtain
\begin{align}
&\left(\frac{\bra{\uparrow\downarrow}-\bra{\downarrow\uparrow}}{\sqrt{2}}\right)\widetilde{B}(\lambda_j)\ket{\uparrow\uparrow}
=\sqrt{2}\sinh(\eta)\sinh(\eta/2)\cosh(\lambda_j)\ , \label{eq:38}\\
&\frac{1}{\sqrt{|q|+1/|q|}}\left(q^{1/2}\bra{\uparrow\downarrow}-q^{-1/2}\bra{\downarrow\uparrow}\right)\widetilde{B}(\lambda_j)\ket{\uparrow\uparrow}
=\frac{\sinh^2(\eta)e^{\lambda_j}}{\sqrt{|q|+1/|q|}}\ , \label{eq:39}
\end{align}
where in Eq. \eqref{eq:39} we have used $q=e^{\eta}$.
\item The tilted N\'eel state 
$\ket{\theta N}=\ket{\theta;\nearrow \swarrow\ldots \nearrow \swarrow }=e^{i\theta/2\sum_j \sigma_j^y}\ket{\uparrow\downarrow\ldots\uparrow\downarrow}$. 
This state is of the form \eqref{eq:12} with $G=2$. The vector $\ket{\varphi}$ is given by
\begin{equation}
\ket{\theta;\nearrow \swarrow}=\sin\left(\frac{\theta}{2}\right)\cos\left(\frac{\theta}{2}\right)\left(\ket{\uparrow\uparrow}-\ket{\downarrow\downarrow}\right)+\cos^2\left(\frac{\theta}{2}\right)\ket{\uparrow\downarrow}-\sin^2\left(\frac{\theta}{2}\right)\ket{\downarrow\uparrow}.
\label{eq:40}
\end{equation}
From Eq. \eqref{eq:40}, we see that in order to write down the recursive formula \eqref{eq:28} for the tilted N\'eel state, in the 
expansion \eqref{eq:23} we have to keep the terms $\ket{\Gamma_0}$, $\ket{\Gamma_1}$ and $\ket{\Gamma_2}$, 
written explicitly in Eqs. \eqref{eq:25}, \eqref{eq:26}, and \eqref{eq:27}. 
We then need the following expressions, which are easily computed
\begin{align}
\braket{\theta;\nearrow \swarrow|\widetilde{B}(\lambda_j)|\uparrow\uparrow}
&=\sinh(\eta) \left[\cos^2\left(\theta/2\right)\sinh(\lambda_j+\eta/2)-\sin^2\left(\theta/2\right)\sinh(\lambda_j-\eta/2)\right]\ , 
\label{eq:41}\\
\braket{\theta;\nearrow \swarrow|\widetilde{B}(\lambda_j)\widetilde{B}(\lambda_l)|\uparrow\uparrow}\ 
&=-\sin\left(\theta/2\right)\cos\left(\theta/2\right)\sinh^2(\eta)\left[\cosh(\eta)\cosh(\lambda_j+\lambda_l)-\cosh(\lambda_j-\lambda_l)\right]. \label{eq:42}
\end{align}
The recursive formula \eqref{eq:28} for the tilted N\'eel can thus be written as 
\begin{equation}
\begin{split}
&S_{M+1}[\theta N](\lambda_1, \ldots ,\lambda_{P})=\\
& =\sin(\theta/2)\cos(\theta/2)\left[ \prod_{k=1}^{P}\sinh^2\left( \lambda_k+\frac{\eta}{2}\right)\right] S_{M}[\theta N](\lambda_1 , \ldots , \lambda_P)+ \\
&+\sum_{j=1}^{P}\sinh^{2M}\left(\lambda_j-\frac{\eta}{2}\right)\braket{\theta;\nearrow \swarrow|\widetilde{B}(\lambda_j)|\uparrow\uparrow} \left[ \prod_{\substack{k=1\\ k\neq j}}^{P}\sinh^2\left( \lambda_k+\frac{\eta}{2}\right) f(\lambda_k,\lambda_j) \right] 
S_{M}[\theta N](\lambda_1 , \ldots , \hat{\lambda}_j , \ldots , \lambda_{P})+\\
&+ \sum_{\substack{l,j\\l<j}}^{P}\sinh^{2M}\left(\lambda_j-\frac{\eta}{2}\right)\sinh^{2M}\left(\lambda_l-\frac{\eta}{2}\right)\left(\prod_{\substack{k=1 \\ k\neq l, j}}^{P} \sinh^2(\lambda_k+\eta/2) f(\lambda_k,\lambda_l)f(\lambda_k,\lambda_j) \right)\times\\
&\hspace{2cm}\times \bra{\theta;\nearrow \swarrow}\widetilde{B}(\lambda_j)\widetilde{B}(\lambda_l)\ket{\widetilde{0}} S_{M}[\theta N](\lambda_1 , \ldots , \hat{\lambda}_l , \ldots , \hat{\lambda}_j, \ldots ,\lambda_{P})\ ,\label{eq:43}
\end{split}
\end{equation}
where $\braket{\theta;\nearrow \swarrow|\widetilde{B}(\lambda_j)|\uparrow\uparrow}$ and $\braket{\theta;\nearrow \swarrow|\widetilde{B}(\lambda_j)\widetilde{B}(\lambda_l)|\uparrow\uparrow}$ are given in Eqs. \eqref{eq:41} and \eqref{eq:42}. 
\item Finally, as an example of a product state of the form \eqref{eq:12} with $G>2$, we consider the ferromagnetic domain state with $G=4$:
\begin{equation}
\ket{FD_4}=\ket{\underbrace{\uparrow\uparrow\downarrow\downarrow}_4\ldots\underbrace{\uparrow\uparrow\downarrow\downarrow}_4}. 
\label{eq:43.1}
\end{equation}
We have now $N=4M$. The only Bethe states with nonzero overlap with the ferromagnetic domain state are those parametrized 
by sets of rapidities with cardinality $2M$. Furthermore, in the expansion \eqref{eq:23} the only term we have to consider is 
$\ket{\Gamma_2}$, the other terms having zero overlap with \eqref{eq:43.1}. 
As usual, we have to compute a matrix element which is easily worked out as 
\begin{equation}
\begin{split}
&\braket{\uparrow\uparrow\downarrow\downarrow|\widetilde{B}(\lambda_j)\widetilde{B}(\lambda_l)|\uparrow\uparrow\uparrow\uparrow}=\sinh^{2}(\lambda_j+\eta/2)\sinh^{2}(\lambda_l+\eta/2)\sinh^{2}(\eta)
\left[\cosh(\eta)\cosh(\lambda_j+\lambda_l)-\cosh(\lambda_j-\lambda_l)\right]\ .
\end{split}
\label{eq:44}
\end{equation}
The recursive formula for the overlap between Bethe states and the ferromagnetic domain state with $G=4$ thus reads
\begin{equation}
\begin{split}
S_{M+1}[FD_4]&(\lambda_1, \ldots ,\lambda_{2(M+1)})=\\
&= \sum_{\substack{l,j\\l<j}}^{2(M+1)}\sinh^{4M}\left(\lambda_j-\frac{\eta}{2}\right)\sinh^{4M}\left(\lambda_l-\frac{\eta}{2}\right)\left(\prod_{\substack{k=1 \\ k\neq l, j}}^{2(M+1)} \sinh^{4}\left(\lambda_k+\frac{\eta}{2}\right)f(\lambda_k,\lambda_l)f(\lambda_k,\lambda_j) \right)\times\\
&\hspace{1cm}\times \braket{\uparrow\uparrow\downarrow\downarrow|\widetilde{B}(\lambda_j)\widetilde{B}(\lambda_l)|\uparrow\uparrow\uparrow\uparrow} S_{M}[FD_4](\lambda_1 , \ldots , \hat{\lambda}_l , \ldots , \hat{\lambda}_j, \ldots ,\lambda_{2(M+1)})\ ,\label{eq:45}
\end{split} 
\end{equation}
where $ \braket{\uparrow\uparrow\downarrow\downarrow|\widetilde{B}(\lambda_j)\widetilde{B}(\lambda_l)|\uparrow\uparrow\uparrow\uparrow}$ 
is given by Eq. \eqref{eq:44}.
\end{enumerate}

\section{Proof of overlaps determinant formulas}\label{determinant_formulas}
In this section we show that the recursive formulas \eqref{eq:36} can be used to prove in a simple way determinant formulas 
for the overlaps between Bethe states and the N\'eel state, the dimer state and the $q$-deformed dimer state, 
firstly derived by B. Pozsgay in Ref. \cite{pozsgay} using Boundary Bethe Ansatz techniques \cite{BABA}. 
Denoting generically with $\ket{\psi}$ the two-site shift invariant state, 
we want to prove the following determinant formula \cite{pozsgay}
\begin{equation}
\begin{split}
\braket{\psi|\lambda_1 , \ldots , \lambda_M} =&
\prod_{j=1}^{M}\left(\frac{\sinh^{2M}\left( \lambda_j -\eta/2 \right)\sinh^{2M}\left( \lambda_j +\eta/2 \right)}{\sinh(2\lambda_j)\sinh(\eta)}\right)\prod_{j=1}^{M}\braket{\varphi|\widetilde{B}(\lambda_j)|\uparrow\uparrow}\times \\
&\times \frac{1}{\prod_{j<k}\sinh(\lambda_j-\lambda_k)\sinh(\lambda_j+\lambda_k)} \det L_M(\lambda_1,\ldots, \lambda_M) \ ,
\label{eq:46}
\end{split}
\end{equation}
where the elements of the $M\times M$ matrix $L$ (not to be confused with the $L$-matrix of the Algebraic Bethe Ansatz) are 
\begin{equation}
\left[L(\lambda_1, \ldots , \lambda_M)\right]_{jk}=\coth^{2j}(\lambda_k-\eta/2)-\coth^{2j}(\lambda_k+\eta/2)\ ,
\label{eq:47}
\end{equation}
and $\braket{\varphi|\widetilde{B}(\lambda_j)|\uparrow\uparrow}$ are given in Eqs. \eqref{eq:37}, \eqref{eq:38}, and \eqref{eq:39} 
for the N\'eel state, the dimer state, and the $q$-deformed dimer state respectively. 
Eq. \eqref{eq:46} is a simple rewriting in our notations of the overlaps formulas in Ref. \cite{pozsgay} for these states. 
From the structure of Eq. \eqref{eq:36} it is natural to look for a solution in terms of the determinant of a certain matrix. 
In fact, recursive formula \eqref{eq:36} has the same form of the Laplace's recursive formula for the determinant of a square matrix.
The proof of Eq. \eqref{eq:46} is indeed straightforward from the recursion relation and it proceeds by induction. 
First, we see that the case $M=1$ is obvious. 
Next, we prove that Eq.  \eqref{eq:46} fulfils recursive formula \eqref{eq:36}. Plugging Eq. \eqref{eq:46} into Eq. \eqref{eq:36} we have
\begin{equation}
\begin{split}
\det  L_{M+1}&(\lambda_1, \ldots ,\lambda_{M+1})= \\
&= \sinh(\eta)\sum_{j=1}^{M+1}\left[ \prod_{\substack{k=1\\ k\neq j}}^{M+1} f(\lambda_k,\lambda_j) \right]
\frac{\sinh(2\lambda_j)\prod_{r=1}^{j-1}\sinh(\lambda_r-\lambda_j)\prod_{r=j+1}^{M+1}\sinh(\lambda_j-\lambda_r)\prod_{r\neq j}^{M+1}\sinh(\lambda_j+\lambda_r)}{\sinh^{2}(\lambda_j-\eta/2)\sinh^{2M+2}(\lambda_j+\eta/2)\prod_{k\neq j}^{M+1}\sinh^2(\lambda_k-\eta/2)} \times \\& \hspace{15mm}\times 
\det L_{M}(\lambda_1 , \ldots ,\widehat{\lambda}_j ,\ldots , \lambda_{M+1})\ ,
\label{eq:48}
\end{split}
\end{equation}
where $f(\mu,\nu)$ is given in Eq. (\ref{eq:3}).
We now define 
\begin{equation}
a_j\equiv \lambda_j+\eta/2, \qquad b_j\equiv \lambda_j - \eta/2,
\label{eq:49}
\end{equation}
and
\begin{equation}
\alpha_k\equiv \text{coth}^2(a_k),\qquad \beta_k \equiv \text{coth}^2(b_k)\ .
\label{eq:50}
\end{equation}
Using the identities
\begin{equation}
\frac{\sinh(\eta) \sinh(2\lambda_j)}{\sinh^2(\lambda_j-\eta/2)\sinh^2(\lambda_j+\eta/2)}=\text{coth}^2(b_j)-\text{coth}^2(a_j)=\beta_j-\alpha_j\ ,
\label{eq:51}
\end{equation}
\begin{equation}
\frac{\sinh(\lambda_j-\lambda_k+\eta) \sinh(\lambda_j+\lambda_k)}{\sinh^2(\lambda_j+\eta/2)\sinh^2(\lambda_k-\eta/2)}=\text{coth}^2(b_k)-\text{coth}^2(a_j)=\beta_k-\alpha_j\ ,
\label{eq:52}
\end{equation}
\begin{equation}
\begin{split}
\left(\prod_{k=1}^{j-1}\sinh(\lambda_k-\lambda_j)\right)&\left(\prod_{r=j+1}^{M+1}\sinh(\lambda_j-\lambda_r)\right)
=(-1)^{j+1}\prod_{\substack{k=1 \\ k\neq j}}^{M+1} \sinh(\lambda_j-\lambda_k)\ ,
\label{eq:53}
\end{split}
\end{equation}
Eq. \eqref{eq:48} can be written as
\begin{equation}
\begin{split}
\det L_{M+1}&(\lambda_1, \ldots , \lambda_{M+1}) = 
\sum_{j=1}^{M+1}(-1)^{j+1}\prod_{k=1}^{M+1}(\beta_k-\alpha_j)\det L_{M}(\lambda_1, \ldots , \widehat{\lambda}_j , \ldots , \lambda_{M+1})\ .
\label{eq:54}
\end{split}
\end{equation}
Consider now Laplace's formula for the determinant of a square matrix $A$ (with elements $a_{jk}$) of size $N$
\begin{equation}
\det A = \sum_{j=1}^{N}(-1)^{N+j}a_{Nj}\det \widetilde{A}_{Nj}\ ,
\label{eq:55}
\end{equation}
where $\widetilde{A}_{Nj}$ is the square matrix of size $N-1$ that results from $A$ by removing the $N$th row and the $j$th column. 
Using Eq. \eqref{eq:55} and the induction hypothesis we see that the r.h.s. of Eq. \eqref{eq:54} is equal to
\begin{equation}
\begin{split}
\det \begin{pmatrix}
  \beta_1-\alpha_1  & \cdots  &\beta_{M+1}-\alpha_{M+1} \\
   \beta_1^2-\alpha_1^2 &\cdots  & \beta_{M+1}^2-\alpha_{M+1}^2 \\
  \vdots & \ddots & \vdots\\
   \beta_1^M-\alpha_1^M& \cdots   & \beta_{M+1}^M-\alpha_{M+1}^M\\[0.6cm]
  (-1)^{M} \prod_{k=1}^{M+1}(\beta_k-\alpha_1)  &\cdots &(-1)^{M} \prod_{k=1}^{M+1}(\beta_k-\alpha_{M+1})
 \end{pmatrix}\ .
\end{split}
 \label{eq:56}
\end{equation}
To conclude the proof it's sufficient to show that \eqref{eq:56} is equal to
\begin{equation}
\det \begin{pmatrix}
  \beta_1-\alpha_1 & \beta_2-\alpha_2  & \cdots  &\beta_{M+1}-\alpha_{M+1} \\
   \beta_1^2-\alpha_1^2&\beta_2^2-\alpha_2^2 &\cdots  & \beta_{M+1}^2-\alpha_{M+1}^2 \\
  \vdots & \vdots & \ddots &\vdots \\
   \beta_1^{M+1}-\alpha_1^{M+1}&\beta_2^{M+1}-\alpha_2^{M+1}  & \cdots & \beta_{M+1}^{M+1}-\alpha_{M+1}^{M+1}
 \end{pmatrix}\ .
 \label{eq:57}
\end{equation}
This can be done as follows. Expand the $j$th entry in the last row of Eq. \eqref{eq:56} as
\begin{equation}
\begin{split}
&(-1)^{M} \prod_{k=1}^{M+1}(\beta_k-\alpha_j)= 
-\alpha_j^{M+1}+\alpha_j^{M}\left(\sum_{i=1}^{M+1}{\beta_i}\right)-\alpha_j^{M-1}\left(\sum_{1 \leq i<j \leq M+1}\beta_i \beta_j\right)+\alpha_j^{M-2}\left(\sum_{i<j<k}\beta_i \beta_j \beta_k\right)+\ldots\ .
\label{eq:58}
\end{split}
\end{equation}
Exploiting the properties of the determinant, we can simplify Eq. \eqref{eq:58} without changing the determinant in Eq. \eqref{eq:56} 
with the following manipulations
\begin{itemize}
\item multiply row $M$ of \eqref{eq:56} by $\left(\sum_{i=1}^{M+1}\beta_i\right)$ and sum it to the row $M+1$;
\item multiply row $(M-1)$ of \eqref{eq:56} by $\left(\sum_{i<j}\beta_i \beta_j\right)$ and subtract it from the row $M+1$;
\item multiply row $(M-2)$ of \eqref{eq:56} by $\left(\sum_{i<j<k}\beta_i \beta_j \beta_k\right)$  and sum it to the row $M+1$.
\item iterate the steps above for the remaining rows of \eqref{eq:56} until the last one.
\end{itemize}
From Eq. \eqref{eq:58} it is easy to see that, at the end of the procedure described above, the last row of the matrix in Eq. 
\eqref{eq:56} is written as
\begin{equation}
\Big( (\beta_{1}^{M+1}-\alpha_1^{M+1}), (\beta_{2}^{M+1}-\alpha_2^{M+1}), \ldots , (\beta_{M+1}^{M+1}-\alpha_{M+1}^{M+1}) \Big)\ ,
\label{eq:59}
\end{equation}
and this concludes the proof.

\section{Conclusions}
\label{conclusions}

In this work we derived a very general recursive formula for the overlaps between Bethe states and a broad class of product states 
in the XXZ spin-$1/2$ chain (which includes all product states considered so far in analytic and numerical computation). 
Explicit examples, i.e. for particular initial states, of recursive formulas are reported in Sec. \ref{examples}.
These recursive formulas are obtained using the Algebraic Bethe Ansatz and rely exclusively on the lattice structure 
of the XXZ model. 
Our approach is straightforwardly generalised to other lattice Bethe Ansatz solvable models such as 
the integrable lattice regularisations of the Lieb-Liniger model \cite{amico, bogoliubov}.

In the case of product states with two-site shift invariance, our recursive formula allows to prove very easily the recently found 
overlaps in Ref. \cite{pozsgay}. As a relevant difference, our proof does not use any concept coming from the Boundary Algebraic  
Bethe Ansatz solution of the classical six-vertex model and it is genuinely based on the solution of the XXZ chain. 
It is then highly desirable to find compact (determinant) solutions of all the recursive formulas listed in Sec. \ref{examples}. 
Of course, a trivial solution of these recursive formulas is given by the formal expression obtained using the Coordinate Bethe Ansatz 
wavefunctions, but such an expression is neither numerically nor analytically tractable. 
Compact solutions would be very useful for  analytical studies of the quench  dynamics of the XXZ model 
based on the representative state approach \cite{ce-13}, on the same lines of  
those presented in Refs. \cite{pozsgay_werner_kormos, wouters_brockmann} for the dimer and N\'eel states respectively.
This would allow to confront with the GGE solution \cite{fcce-13} for a wide class of states.

It is worth stressing that our recursive formulas refer to off-shell scalar products, that is the rapidities $\{\lambda_j\}$ 
defining the Bethe states do not necessarily fulfil the Bethe equations \eqref{eq:9}. 
Provided that compact solutions for recursive formulas can be found, following Ref. \cite{brockmann_I} for  two-site shift invariant states,
they could be used as a starting point for obtaining simplified, on-shell formulas suitable 
for taking the thermodynamic limit. 
The question remains whether compact formulas exist in the XXZ model for the overlaps between 
Bethe states and other states besides those discussed in section \ref{determinant_formulas}.
We mention that product states of the form (\ref{eq:12}) can be used as good approximations of the ground states of 
gapped Hamiltonian with correlation length $\xi\alt G$ in a matrix product state representation, on the same lines 
of what done for the GGE in Ref. \cite{fcce-13}. 

\section*{Acknowledgments}   
PC acknowledge the ERC  for financial  support under  Starting Grant 279391 EDEQS.

\end{document}